\documentclass[twocolumn,showpacs,preprintnumbers,amsmath,amssymb]{revtex4}
\usepackage{graphicx}% Include figure files
\usepackage{dcolumn}% Align table columns on decimal point
\usepackage{bm}% bold math

\begin{document}

\preprint{APS/123-QED}

\title{Inter-Band Effects of Magnetic Field on Hall Conductivity in Multi Layered Massless Dirac Fermion System $\alpha$-(BEDT-TTF)$_2$I$_3$ }% Force line breaks with \\

\author{Naoya Tajima}
\email{naoya.tajima@sci.toho-u.ac.jp}
\author{Reizo Kato$^1$}%
\author{Shigeharu Sugawara$^2$}
\author{Yutaka Nishio}%
\author{Koji Kajita}%
\affiliation{%
Department of Physics, Toho University, Miyama 2-2-1, Funabashi-shi, Chiba JP-274-8510, Japan \\
$^1$RIKEN, Hirosawa 2-1, Wako-shi, Saitama JP-351-0198, Japan\\
$^2$Department of Physics, Faculty of Science and Technology, Tokyo University of Science, Noda, Chiba JP-278-8510, Japan}%

\date{\today}% It is always \today, today,

%%%%%%%%%%%%%%%%%%%%%%%%%%%%%%%%%%%%%%%%%%%%%%%%%%%%%%%%%%%%%%%%%%%%%%%%%%%%%%%%%%%%%%%%%
\begin{abstract}
We have discovered two-dimensional zero-gap material with a layered structure in the organic conductor $\alpha$-(BEDT-TTF)$_2$I$_3$ under high hydrostatic pressure. In contrast to graphene, the electron-hole symmetry is not good except at the vicinity of the Dirac points. Thus, temperature dependence of the chemical potential, $\mu$, plays an important role in the transport in this system. The experimental formula of $\mu$ is revealed. We succeeded in detecting the inter-band effects of a magnetic field on the Hall conductivity when $\mu$ passes the Dirac point.
\end{abstract}


\maketitle

%-------------improve
Since Novoselov {\it et al.} \cite{rf:1} and Zhang {\it et al.} \cite{rf:2} experimentally demonstrated that graphene is a zero-gap system with massless Dirac particles, such systems have fascinated physicists as a source of exotic systems and/or new physics.
%--------------------
The Dirac fermion system, on the other hand, was also realized in the quasi-two dimensional (2D) organic conductor $\alpha$-(BEDT-TTF)$_2$I$_3$ (inset of Fig. 1(b)) under high pressures.\cite{rf:3,rf:4,rf:5,rf:6, rf:7} In contrast to graphene, this is the first bulk material with a zero-gap energy band. One of the characteristic features of the Dirac fermion system is seen in the magnetic field normal to the conductive layer. In the magnetic field, the energy of Landau levels (LLs) in zero-gap systems is expressed as $E_{\rm nLL}=\pm \sqrt{2e \hbar v_{\rm F}^2 |n||B|}$, where $v_{\rm F}$ is the Fermi velocity, $n$ is the Landau index, and $B$ is the magnetic field strength. One important difference between zero-gap conductors and conventional conductors is the appearance of a (n=0) LL at zero energy.\cite{rf:8} This special LL is called the zero-mode LL. The characteristic features of zero-mode Landau carriers, including spin splitting, are clearly seen in transport \cite{rf:9, rf:10, rf:11, rf:12} and specific heat phenomena.\cite{rf:13} 

A weak magnetic field, on the other hand, also gives us characteristic phenomena. According to the theory of Fukuyama, the vector potential plays an important role in inter-band excitation in electronic systems with a vanishing or narrow energy gap.\cite{rf:14} The orbital movement of virtual electron-hole pairs gives rise to anomalous orbital diamagnetism and the Hall effect in a weak magnetic field.\cite{rf:14} These are called the interband effects of the magnetic field. This discovery inspired us to examine the interband effects of the magnetic field in $\alpha$-(BEDT-TTF)$_2$I$_3$. In this paper, we demonstrate that these effects give rise to anomalous Hall conductivity in $\alpha$-(BEDT-TTF)$_2$I$_3$. Moreover, to detect these effects, we succeeded in developing an experimental formula for the temperature dependence of the chemical potential, $\mu$. Bismuth and graphite are the most well-known materials that serve as testing ground for the interband effects of magnetic fields.\cite{rf:15, rf:16} To our knowledge, however, $\alpha$-(BEDT-TTF)$_2$I$_3$ is the first organic material in which the inter-band effects of the magnetic field have been detected. 

A crystal of $\alpha$-(BEDT-TTF)$_2$I$_3$ consists of conductive layers of BEDT-TTF molecules and insulating layers of I$_3^-$ anions as shown in the inset of Fig. 1(b).\cite{rf:17} As the conductive layers are separated by the insulating layers, carriers in this system have a strong two-dimensional nature. 
%--------------improve
Two dimensional electron systems on conducting layers couple with each other by sufficiently weak interlayer tunneling to form a layered quasi-2D conductor. The ratio of the in-plane conductivity to the out-of-plane conductivity is more than 1000.
%---------------------
Under ambient pressure, it behaves as a metal down to 135 K at which point it undergoes a phase transition to a charge-ordered insulator. In the low-temperature phase, a horizontal-charge-stripe pattern for +1 $e$ and 0 is formed.\cite{rf:18,rf:19,rf:20,rf:21} When a crystal is exposed to a high hydrostatic pressure above 1.5 GPa, the metal-insulator transition is suppressed, and the Dirac-fermion system is realized stably.\cite{rf:3, rf:4, rf:5} 
%---------------improve
The suppression of the metal-insulator transition by a pressure above 1.5 GPa is accompanied by the disappearance of the charge-ordering state as shown by Raman experiment.\cite{rf:21} Above 1.5 GPa, the resistivity is independent of pressure.\cite{rf:5} Dirac fermion system remains stable at least up to 2.0 GPa. 
%---------------------

Realistic theory predicts that the interband effects of the magnetic field are detected by measuring the Hall conductivity $\sigma_{xy}$ or the magnetic susceptibility at the vicinity of the Dirac points (crossing points of the zero-gap structure).\cite{rf:16, rf:22} In order to detect these effects, we should control $\mu$. In this material, however, the multilayered structure makes control of $\mu$ by the field-effect-transistor method much more difficult than in the case of graphene. Hence, we present the following idea. 

We find two types of samples in which electrons or holes are slightly doped by unstable I$_3^-$ anions. The doping gives rise to strong sample dependences of the resistivity or the Hall coefficient at low temperatures (Figs. 1 and 2). In particular, the sample dependence of the Hall coefficient $R_{\rm H}$ is intense. In the hole-doped sample as shown in the inset of the lower part of Fig. 2(a), $R_{\rm H}$ is positive over the whole temperature range (Fig. 2(a)). In the electron-doped sample as shown in the inset of Fig. 2(b), however, the polarity is changed at low temperatures (Fig. 2(b)). The change in polarity of $R_{\rm H}$ is understood as follows. In contrast to graphene, the present electron-hole symmetry is not good except at the vicinity of the Dirac points.\cite{rf:23} Thus, $\mu$ must be dependent on temperature. Of significance is the fact that according to the theory by Kobayashi {\it et al.}, when $\mu$ passes the Dirac point ($\mu=0$), $R_{\rm H}=0$.\cite{rf:22} Thus, $R_{\rm H}$ at the vicinity of $R_{\rm H}=0$ for electron-doped samples must be determined to detect the inter-band effects of the magnetic field. 

%-------------------improve
Here we adduce other examples for the effect of unstable I$_3^-$ anions. It was also seen in the superconducting transition of the organic superconductors $\beta$-(BEDT-TTF)$_2$I$_3$ (Ref. 24) and $\theta$-(BEDT-TTF)$_2$I$_3$.\cite{rf:25}
%--------------------------

%-------------------improve
Single crystals of $\alpha$-(BEDT-TTF)$_2$I$_3$ were synthesized by the electrolysis method.\cite{rf:17} The typical size of a crystal was $0.8 \times 0.4 \times 0.04$ mm$^3$.
%--------------------------
The resistivities and the Hall coefficients of seven samples at $p=1.8$ GPa were measured in magnetic fields of up to 0.01 T at temperatures below 20 K. Experiments were conducted as follows. A sample to which six electrical leads were attached was put in a Teflon capsule filled with a pressure medium (Idemitsu DN-oil 7373), and then the capsule was set in a clamp-type pressure cell made of the hard alloy MP35N. The pressure loss at low temperatures was less than 0.15 GPa. Resistance was measured by the conventional DC method with six probes as shown in the inset of Fig. 2(a). An electrical current of 10 $\mu$A was applied in the 2D plane along the $a$ axis. Magnetic fields were applied in the direction normal to the 2D plane. 
%-------------------improve
The doping levels were estimated to be 1-10 ppm for seven samples from the Hall coefficient at lowest temperature. This result supports the theory of Kobayashi {\it et al}. Note that the difference in the sample was not able to be distinguished by appearance. 
%--------------------------

%------------------------------------------------------------------------------
%------------- Figure 1 -------------------------------------------------------
%------------------------------------------------------------------------------
\begin{figure}
\includegraphics[viewport = 0 200 550 650, scale=.4, clip]{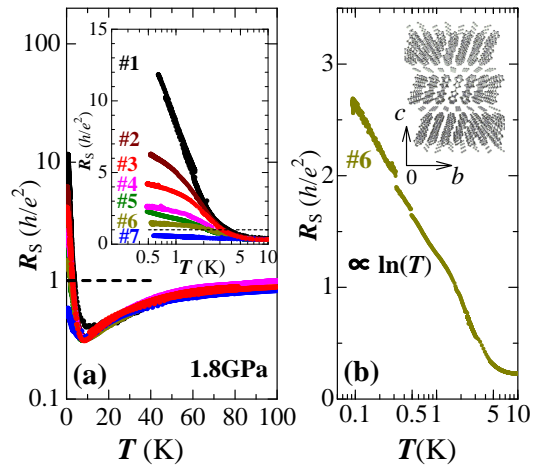}
\caption{\label{fig:1}  (Color online) (a) Temperature dependence of $R_{\rm S}$ for seven samples under a pressure of 1.8 GPa. The inset $R_{\rm S}$ at temperatures below 10 K. (b) Temperature dependence of $R_{\rm S}$ for sample 6. It was examined down to 80 mK. The inset shows the crystal structure of $\alpha$-(BEDT-TTF)$_2$I$_3$ viewed from the $a$ axis.}
\end{figure}
%------------------------------------------------------------------------------
%------------------------------------------------------------------------------
%------------------------------------------------------------------------------

Figure 1 shows the temperature dependence of the resistivity $\rho$ per layer (sheet resistance, $R_{\text{S}}=\rho /C$, where $C=1.75$ nm is the lattice constant along the direction normal to the 2D plane). Note that the concept of \lq\lq resistivity per layer\rq\rq applies to this system in which each conductive layer contributes to the transport almost independently. In this figure, we see that the sheet resistance depends on the temperature very weakly except for the region below 7 K, at which an increase in resistance is observed. It varies from a value of approximately seven times the quantum resistance $h/e^2$=25.8 k$\Omega$ at 100 K to approximately twice $h/e^2$ at 7 K. Then, the reproducibility of the data is checked in six samples. Above 7 K, we find only weak sample dependence. This sheet resistance is understood as follows. 

Carrier density, $n_c$, written as $n_c \propto T^2$, is a characteristic feature of 2D zero-gap conductors.\cite{rf:4, rf:5} Thus, the Hall coefficient is proportional to $T^{-2}$ as shown in Fig. 2(b). Carrier mobility, on the other hand, is determined as follows. According to Mott's argument \cite{rf:26}, the mean free path $l$ of a carrier subjected to elastic scattering can never be shorter than the wavelength $\lambda$ of the carrier, so $l \geq \lambda$. For the cases in which scattering centers exist at high densities, $l \sim \lambda$. As the temperature is decreased, $l$ becomes long because $\lambda$ becomes long with the decreasing energy of the carriers. As a result, the mobility of carriers increases in proportion to $T^{-2}$ in the 2D zero-gap system. Consequently, sheet resistance is given by $R_{\rm S} =  h/e^2$, which is independent of temperature.

%------------------------------------------------------------------------------
%------------- Figure 2 -------------------------------------------------------
%------------------------------------------------------------------------------
\begin{figure}
\includegraphics[viewport = 0 0 600 450, scale=.4, clip]{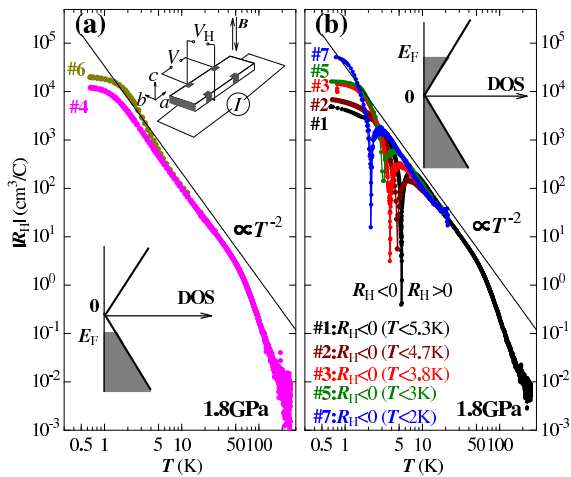}
\caption{\label{fig:2} (Color online) Temperature dependence of the Hall coefficient for (a) hole-doped-type and (b) electron-doped-type samples. Note that in this figure, the absolute value of $R_{\rm H}$ is plotted. Thus, the dips in (b) indicate a change in the polarity. The inset in the upper part of (a) shows the configuration of the six electrical contacts. The schematic illustration of the Fermi levels for the hole-doped type and the electron-doped type are shown in the inset of the lower part of (a) and the inset of (b), respectively.
} 
\end{figure}
%------------------------------------------------------------------------------
%------------------------------------------------------------------------------
%------------------------------------------------------------------------------

Below 7 K, the sample dependence due to the effects of unstable I$_3^-$ anions appears strongly. 
%----------------improve
A rise of $R_{\rm S}$ may be a symptom of localization because it is proportional to $\log T$ as shown in the inset of Fig. 1(a). The unstable I$_3^-$ gives rise to a partially incommensurate structure in the BEDT-TTF layers. As for sample 6, the $\log T$ law of $R_{\rm S}$ was examined down to 80 mK. The $\log T$ law of resistivity, on the other hand, is also characteristic of the transport of Kondo-effect systems. In graphene, recently, Chen {\it et al.} have demonstrated that the interaction between the vacancies and the electrons give rise to Kondo-effect systems.\cite{rf:27} The origin of the magnetic moment was the vacancy. In $\alpha$-(BEDT-TTF)$_2$I$_3$, on the other hand, Kanoda {\it et al.} detected anomalous NMR signals which could not be understood based on the picture of Dirac fermion systems at low temperatures.\cite{rf:28} 
%In this material, the partially incommensurate structure of BEDT-TTF layers may produce the magnetic moment. 
We do not yet know which in the localization, the Kondo-effect or other mechanisms, are the answer for the origin of the $\log T$ law of $R_{\rm S}$. This answer, however, will reply to the question of why the sample with the higher carrier density (lower $R_{\rm H}$ saturation value in Fig. 2) exhibits the higher $R_{\rm S}$ at low temperature. Further investigation should lead us to interesting phenomena.
%-----------------------

Another impressive phenomenon is seen in $R_{\rm H}$, as shown in Fig. 2(b). The polarity of $R_{\rm H}$ at low temperature indicates which was doped: electrons or holes. The saturation value at the lowest temperature, on the other hand, depends on the doping density, $n_{\rm d}$, as $n_{\rm d}=n_{\rm S}/C=1/e/R_{\rm H}$, where $n_{\rm S}$ is the sheet density. In this work, to detect the inter-band effects of the magnetic field, we focused on the behavior of $R_{\rm H}$ in which the polarity is changed. 

The first step is to examine the temperature dependence of $\mu$. As mentioned before, we believe $\mu$ passes the Dirac point at the temperature $T_0$, shown as $R_{\rm H}=0$.\cite{rf:22} The sheet density $n_{\rm S}$, on the other hand, is approximately proportional to $T_0^2$ as shown in Fig. 3(a). This result suggests that $\mu$ is to be written as $\mu/k_{\rm B}=E_{\rm F}/k_{\rm B}-AT=0$ at $T_0$ approximately because $E_{\rm F} \propto \sqrt{n_{\rm S}}$, where $A$ is the fitting parameter depending on the electron-hole symmetry ($v_{\rm F}^{\rm h}/v_{\rm F}^{\rm e}$). Thus, we obtain the $E_{\rm F}$ versus $T_0$ curve in Fig. 3(b). $E_{\rm F}$ is estimated from the relationship $n_{\rm S}=E_{\rm F}^2/4\pi \bar{v}_{\rm F}^2$, where the averaged Fermi velocity $\bar{v}_{\rm F}$, is estimated to be approximately $3.3 \times 10^4$ m/s from the interlayer magnetoresistance.\cite{rf:10} Note that the weak sample dependence of both $R_{\rm S}$ and $R_{\rm H}$ at temperatures above 7 K (Figs. 1 and 2) strongly indicates that the $v_{\rm F}$ values of all samples are almost the same. When we assume that $A$ is independent of $E_{\rm F}$, $A$ is estimated to be approximately 0.24 from Fig. 3(b). Thus, we examine the temperature dependence of $\mu$ as $\mu/k_{\rm B}=E_{\rm F}/k_{\rm B}-AT$ with $A \sim 0.24$. This experimental formula reproduces well the realistic theoretical curve by Kobayashi {\it et al.},\cite{rf:22} as shown in Fig. 3(c). Our simple calculations, on the other hand, also reproduce well this curve when we assume $v_{\rm F}^{\rm h}/v_{\rm F}^{\rm e} \sim 1.2$, where $v_{\rm F}^{\rm h}$ and $v_{\rm F}^{\rm e}$ are the holes and electrons of the Fermi velocity, respectively. This is the electron-hole symmetry in our system.  

%------------------------------------------------------------------------------
%------------- Figure 3 -------------------------------------------------------
%------------------------------------------------------------------------------
\begin{figure}
\includegraphics[viewport = 0 00 600 450, scale=.4, clip]{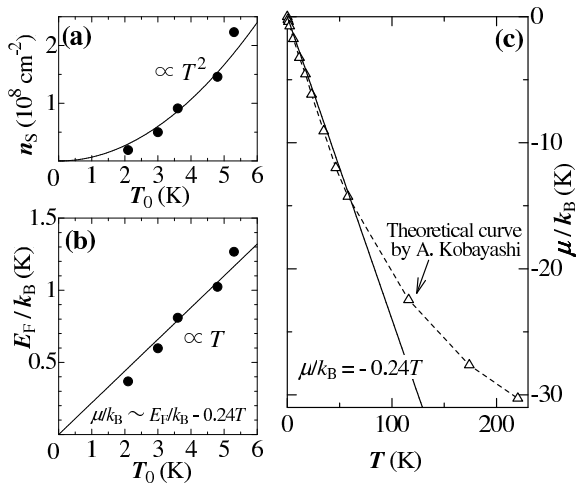}
\caption{\label{fig:3}
(a) Sheet electron density $n_{\rm S}$ for five samples plotted against temperature at $R_{\rm H}=0$. (b) $E_{\rm F}$ was estimated from the relationship $n_{\rm S}= E_{\rm F}^2/(4\pi \hbar^2 v_{\rm F}^2)$ with $v_{\rm F} \sim 3\times 10^4$ m/s. From this curve, the temperature dependence of $\mu$ is written approximately as $\mu /k_{\rm B} = E_{\rm F}/k_{\rm B}-AT$ with $A \sim 0.24$ when we assume $A$ is independent of $E_{\rm F}$. (c) Temperature dependence of $\mu$ for $E_{\rm F}=0$. Our experimental formula is quantitatively consistent with the theoretical curve of Kobayashi {\it et al.} \cite{rf:22} at temperature below 100 K. 
}
\end{figure}
%------------------------------------------------------------------------------
%------------------------------------------------------------------------------
%------------------------------------------------------------------------------

The second step is to calculate the Hall conductivity as $\sigma_{xy}=\rho_{yx}/(\rho_{xx}\rho_{yy}+\rho_{yx}^2)$. In this calculation, we assume $\rho_{xx}=\rho_{yy}$ for the following reasons. According to band calculation, the energy contour of the Dirac cone is highly anisotropic.\cite{rf:7, rf:23} In the galvano-magnetic phenomena, however, the anisotropy is averaged and the system looks very much isotropic. A simple calculation indicates that the variation in the mobility with the current direction is within a factor of 2. Experimentally, Iimori {\it et al.} showed that the anisotropy in the in-plane conductivity is less than 2.\cite{rf:29}

%------------------------------------------------------------------------------
%------------- Figure 4 -------------------------------------------------------
%------------------------------------------------------------------------------
\begin{figure}
\includegraphics[viewport = 0 0 600 430, scale=.4, clip]{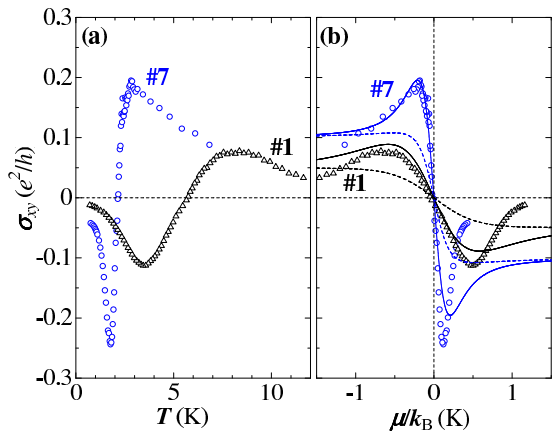}
\caption{\label{fig:4}  (Color online) (a) Temperature dependence of the Hall conductivity for samples 1 and 7. (b) Chemical-potential dependence of the Hall conductivity for samples 1 and 7. Solid lines and dashed lines are the theoretical curves with and without the interband effects of the magnetic field by Kobayashi {\it et al.}, respectively.\cite{rf:22}
}
\end{figure}
%------------------------------------------------------------------------------
%------------------------------------------------------------------------------
%------------------------------------------------------------------------------

Based on this assumption, we show the temperature dependence of $\sigma_{xy}$ for samples 1 and 7 in Fig. 4(a) as an example. We see the peak structure in $\sigma_{xy}$ at the vicinity of $\sigma_{xy} =0$. This peak structure is the anomalous Hall conductivity originating from the interband effects of the magnetic field. The realistic theory indicates that the Hall conductivity without the interband effects of the magnetic field has no peak structure \cite{rf:22}.

In the last step, we redraw $\sigma_{xy}$ in Fig. 4(b) as a function of $\mu$ using the experimental formula, $\mu=E_{\rm F}-AT$ with $A=0.24$. It should be compared with the theoretical curve,\cite{rf:22} $\sigma_{xy}^{theory}$ at $T=0$. Our experimental data are roughly expressed as $\sigma_{xy} \sim g \sigma_{xy}^{theory}$, where $g$ is a parameter that depends on temperature because the effect of thermal energy on the Hall effect is strong. Note that $\sigma_{xy}$ depends strongly on temperature. The energy between two peaks, is the damping energy that depends on the density of scattering centers in a crystal. The intensity of the peak, on the other hand, depends on the damping energy and the tilt of the Dirac cones.\cite{rf:22} 

Lastly, we briefly mention the zero-gap structure in this material. The smooth change in the polarity of $\sigma_{xy}$ is also evidence that this material is an intrinsic zero-gap conductor. Nakamura demonstrated theoretically that in a system with a finite energy gap, $\sigma_{xy}$ is changed in a stepwise manner.\cite{rf:30} 

In conclusion, $\alpha$-(BEDT-TTF)$_2$I$_3$ under high hydrostatic pressure is an intrinsic zero-gap conductor. In contrast to graphene, the electron-hole symmetry is not good except at the vicinity of the Dirac points. Thus, the temperature dependence of $\mu$ plays an important role in the transport in this system. We revealed  that $\mu$ depends on temperature as $\mu/k_{\rm B}=E_{\rm F}/k_{\rm B}-AT$ with $A \sim 0.24$. We succeeded in detecting the interband effects of the magnetic field on the Hall conductivity when $\mu$ passes the Dirac point. Good agreement between experiment and theory was obtained. This system offers a testing ground for a new type of particles, namely, massless Dirac fermions with a layered structure and anisotropic Fermi velocity.

%%%%%%%% Acknowleagements %%%%%%%%%%%%%%%%%%%%%%%%%%%%%%%%%%%%%%%%%%%%%%%%%%%%%%%%%%%%%%%
We thank Prof. T. Osada, Prof. A. Kobayashi, Dr. S. Katayama, Prof. Y. Suzumura, Dr. R. Kondo, Dr. T. Morinari, Prof. T. Tohyama, and Prof. H. Fukuyama for valuable discussions. This work was supported by Grants-in-Aid for Scientific Research (No. 22540379 and No.22224006) from the Ministry of Education, Culture, Sports, Science and Technology, Japan.

%%%%%%%% References %%%%%%%%%%%%%%%%%%%%%%%%%%%%%%%%%%%%%%%%%%%%%%%%%%%%%%%%%%%%%%%%%%%%%
%\bibliography{transport}

%%%%%%%%%%%%%%%%%%%%%%%%%%%%%%%%%%%%%%%%%%%%%%%%%%%%%%%%%%%%%%%%%%%%%%%%%%%%%%%%%%%%%%%%%
\end{document}